\documentclass[aps,prb,groupedaddress,amsmath,amssymb]{revtex4}

\usepackage{graphicx}
\usepackage{dcolumn}

\newcommand {\TaGex}{Ta$_{\, \textrm{x}}$Ge$_{\, 1-\textrm{x}}$/Ge\ }%
\newcommand {\grad}{^{\circ}}%
\newcommand {\eq}{\begin{equation}}%
\newcommand {\qe}{\end{equation}}%
\newcommand {\Tg}{T_{g}}%

\begin{document}

\title{Vortex dynamics and states of artificially layered superconducting films with correlated defects}

\author{Andreas Engel}
\email{andreas.engel@dlr.de}
\altaffiliation{present address: DLR, Institute of Space Sensor Technology and Planetary Exploration, Rutherfordstr. 2, 12105 Berlin, Germany}
\author{H.~J.~Trodahl}
\affiliation{School of Chemical and Physical Sciences, Victoria University of Wellington, Wellington, New Zealand}

\date{\today}

\begin{abstract}
Linear resistances and $IV$-characteristics have been measured over a wide range in the parameter space of the mixed phase of multilayered a-\TaGex films. Three films with varying interlayer coupling and correlated defects oriented at an angle $\approx 25\grad$ from the film normal were investigated. Experimental data were analyzed within vortex glass models and a second order phase transition from a resistive vortex liquid to a pinned glass phase. Various vortex phases including changes from three to two dimensional behavior depending on anisotropy have been identified. Careful analysis of $IV$-characteristics in the glass phases revealed a distinctive $T$ and $H$-dependence of the glass exponent $\mu$. The vortex dynamics in the Bose-glass phase does not follow the predicted behavior for excitations of vortex kinks or loops.
\end{abstract}


\maketitle

\section{Introduction}

The early recognition that the most interesting cuprate superconductors consist of alternating layers of superconducting and insulating material has encouraged an enormous effort to model the complex vortex structure and dynamics in such layered systems. \cite{Blatter94} Difficulties achieving strong pinning of vortices in these layered systems, especially those in which the Josephson coupling between adjacent superconducting layers is weak, lends special urgency to the problem, for weakly pinned vortices lead to resistive losses. The research in the cuprates has been carried out on several fronts, with a search for enhanced pinning in these layered systems running parallel to advances in the modelling and understanding of the thermodynamic vortex states in material with various pinning properties. Thus the former work has concentrated on the introduction of correlated defects which traverse many layers, while more fundamental work has demonstrated a first-order solid-liquid vortex transition in samples which are largely free of pinning defects. \cite{Schilling96,Roulin96} The less dramatic phase transition in a pinned vortex fluid is usually regarded and treated as second order, \cite{Fisher89,Fisher91,Nelson93} though there continue to be doubts as to whether there is a thermodynamic phase transition at all in this situation. \cite{Strachan01,Landau02}

It is fully recognized that the state and dynamics of vortices in these material depend strongly on the degree of coupling between the superconducting layers, so that, for example, vortex motion is much more easily controlled in YBa$_{2}$Cu$_{3}$O$_{7-\delta}$ (Y123) than in the less strongly coupled layers in Bi$_{2}$Sr$_{2}$CaCu$_{2}$O$_{8+\delta}$ (Bi2212). There are other examples of cuprates in which the anisotropy and coupling is intermediate between these two, but the lack of good single crystals spanning a full range of anisotropies has encouraged a focus almost exclusively on these two systems. The range of correlated defects is similarly limited, and has depended largely on heavy ion damage, and to a lesser extent on grain boundaries. Although we now understand much of the vortices in the materials, it is important to test that understanding in a greater range of layering characteristics and defect geometries. 

One option available for a fuller study is based on the preparation of artificially layered superconductors. There have indeed been a number of studies on layered films, including multilayers of Nb/Ge, \cite{Ruggiero80,Ruggiero82,Koorevaar94} Mo/Ge, \cite{White91} and Y123/PrBa$_{2}$Cu$_{3}$O$_{7}$. \cite{Brunner91,Triscone94,Radovan99,Yang99} For some years we have contributed to the work with experiments performed on Ta$_{0.3}$Ge$_{0.7}$/Ge multilayers. The system is attractive on two important counts: (i) the constituents form a consistent amorphous structure across the entire range of thicknesses required for the study, and (ii) all of the important measurements are at temperatures below $2.17$~K. The first of these assures that the layers do not change their properties as their thickness is varied, and furthermore encourages the formation of smooth layers. The second allows measurements to be performed while the films are immersed in superfluid He; the excellent heat extraction ensures a temperature very close to the bath temperature even at the high currents required to drive the material into the normal state.

In a number of earlier papers we have described the behavior of homogeneous TaGe films, \cite{Ruck97} of multilayers with varying coupling strengths, \cite{Trodahl96,Ruck97} and of non-layered films with correlated defects. \cite{Abele99} More recently we reported some aspects of the behavior of a film in which correlated defects were introduced into strongly coupled layered films. \cite{Engel01} The present paper completes the picture as regards the effects of correlated defects in a layered environment, and seeks to answer specific questions in relation to the existence of a second order phase transition for varying interlayer coupling strengths and the vortex dynamics in the solid vortex state in the presence of correlated disorder.

The study we report is based on thin films prepared by vacuum deposition. Alternating layers are formed by simply alternating the composition of the vapor stream, and correlated defects threading the films at about $20$--$30\grad$ to the normal are introduced by depositing onto a substrate held at an oblique angle to the vapor stream. The vortex dynamics are then investigated by $I$/$V$ measurements, for vortices aligned both along the extended defects and at a large angle to them. The measurements are performed at many magnetic fields and temperatures in order to sample the full phase diagram.

The remainder of this paper is organized as follows: We start in Sec.~\ref{expdetails} with experimental housekeeping followed in Sec.~\ref{angular} with an investigation of $R$ vs.\ $\theta$ to demonstrate the presence of correlated defects and to establish their orientation. In Sec.~\ref{taff} we present measurements of $R$ vs.\ $T$ (below $H_{c2}$ but far from the solid vortex state) to estimate activation energies, followed by $IV$-characteristics (\ref{ivcurves}) to establish phase diagrams (\ref{phasediagrams}), and detailed investigations of the dynamics in the glass phase (\ref{dynamics}). In each of the last four of these there will be two subheadings contrasting strongly to weakly coupled samples. Finally (\ref{conclusions}) the conclusions.

\section{Samples and Experimental Details}\label{expdetails}

The amorphous multilayered films in this study were prepared by vapor deposition onto glass substrates in a vacuum chamber with a base pressure of less than $10^{-8}$~mbar. Ta was evaporated from an electron-gun heated source, and Ge from a tungsten boat. The two evaporation rates were independently controlled to within $5$\% of their average with quartz crystal deposition monitors. A Ta concentration of about $35$\% was chosen to maximize $T_{c}$. \cite{Trodahl96} The layering characteristics were achieved by periodically placing a shutter across the Ta vapor stream. Tilting the substrates with respect to the incoming vapor flux resulted in the formation of columns with regions of lower density in between them. \cite{Kranenburg94} In previous studies we have demonstrated the effectiveness of these regions as extended pinning sites. \cite{Abele99,Engel01} The layering and defect parameters were established by transmission electron microscopy (TEM), and their chemical composition has been measured by Rutherford backscattering spectroscopy (RBS). \cite{Engel01a} Table~\ref{table1} lists these details for the films contributing to this report.
\begin{table}
\caption{Structural parameters of the films in this report. After the sample identifiers the second column gives the superconducting (s) and insulating (i) layer thicknesses [nm]. The Ta-concentrations of the Ta$_{\, \textrm{x}}$Ge$_{\, 1-\textrm{x}}$ superconducting layers are given in percent in the third column and the last column lists the orientation of the defect structure with respect to the film normal as determined from resistance measurements. The defect angles determined from TEM photographs are $1\grad$ to $4\grad$ smaller. \label{table1}}
\begin{ruledtabular}
\begin{tabular}{lrrr}
 & \multicolumn{1}{c}{s/i} & \multicolumn{1}{c}{Ta} & \multicolumn{1}{c}{$\beta$} \\\hline
C40 & $13.6$/$3.5$ & $32\%$ & $23.8\grad$ \\
J40 & $8.8$/$3.1$ & $35\%$ & $29.3\grad$ \\
P40 & $13.0$/$7.5$ & $37\%$ & $26.3\grad$ \\
\end{tabular}
\end{ruledtabular}
\end{table}

All three films for which we report results here were deposited at an angle of $40\grad$ to the substrate normal, leading to columns oriented near $25\grad$ to the normal. We have previously demonstrated that films deposited at this angle show significantly enhanced pinning, especially for magnetic fields well aligned with the columnar orientation. \cite{Engel01} Two of the films, labelled as C40 and J40, show relatively strong Josephson coupling between adjacent layers, due to their Ge layers being well below the $5$~nm thickness above which that coupling becomes weak. Note that J40 has superconducting TaGe layers that are about $30$\% thinner than C40, which we show below does not affect the qualitative behavior of their superconductivity. The third film, P40, has Ge layers well above $5$~nm, ensuring that the coupling is weak. It will be demonstrated below that the strength of this coupling has a profound effect on the vortex hopping energy, the phase diagram and the vortex dynamics in the vortex glass phase.

\begin{figure}
\includegraphics[width=85mm,keepaspectratio]{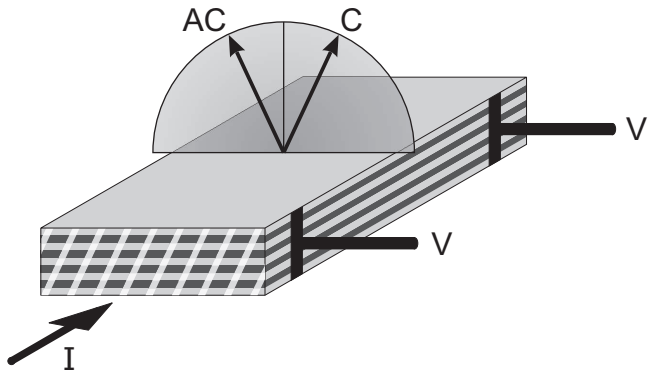}
\caption{Schematic drawing of the measurement geometry, showing the applied current parallel to the film and perpendicular to the columnar microstructure and applied magnetic field. The field could be rotated in the plane perpendicular to the applied current; the C-direction (parallel to the microstructure) and the symmetric AC-direction (large angles to the structure) are indicated. The terminals for measuring the voltage drop along the bridge are also shown. \label{figure0}}
\end{figure}

The films were prepared for the four-terminal $I/V$ measurements by scribing a path $2$~mm wide and $5$~mm long, defining the current to flow perpendicular to the columnar microstructure (Fig.~\ref{figure0}). %
They were inserted in the cryostat in an orientation which allowed the magnetic field to be applied at any arbitrary direction in the plane perpendicular to the current, including along the defect structure. Below we signify that geometry, with the field parallel to the extended coplanar defects, by the letter C. As discussed above the columns are directed approximately $25\grad$ from the normal, so that the symmetric orientation relative to the normal, labelled here the AC direction, placed the field about $50\grad$ to the columns. We will compare measurements made in these two field directions below.

The samples were immersed directly in $^{4}$He, with a magnetic field applied by either a Varian electromagnet or a superconducting solenoid capable of reaching $7$~T. The software-controlled current was supplied by a Keithley model 224 precision current source and the voltage measured with a Keithley model 182 nanovoltmeter. More complete details of the measurements systems are found elsewhere. \cite{Engel01,Engel01b} 

\section{Angular Dependence of Resistance}\label{angular}

\begin{figure}
\includegraphics[angle=-90,width=85mm,keepaspectratio]{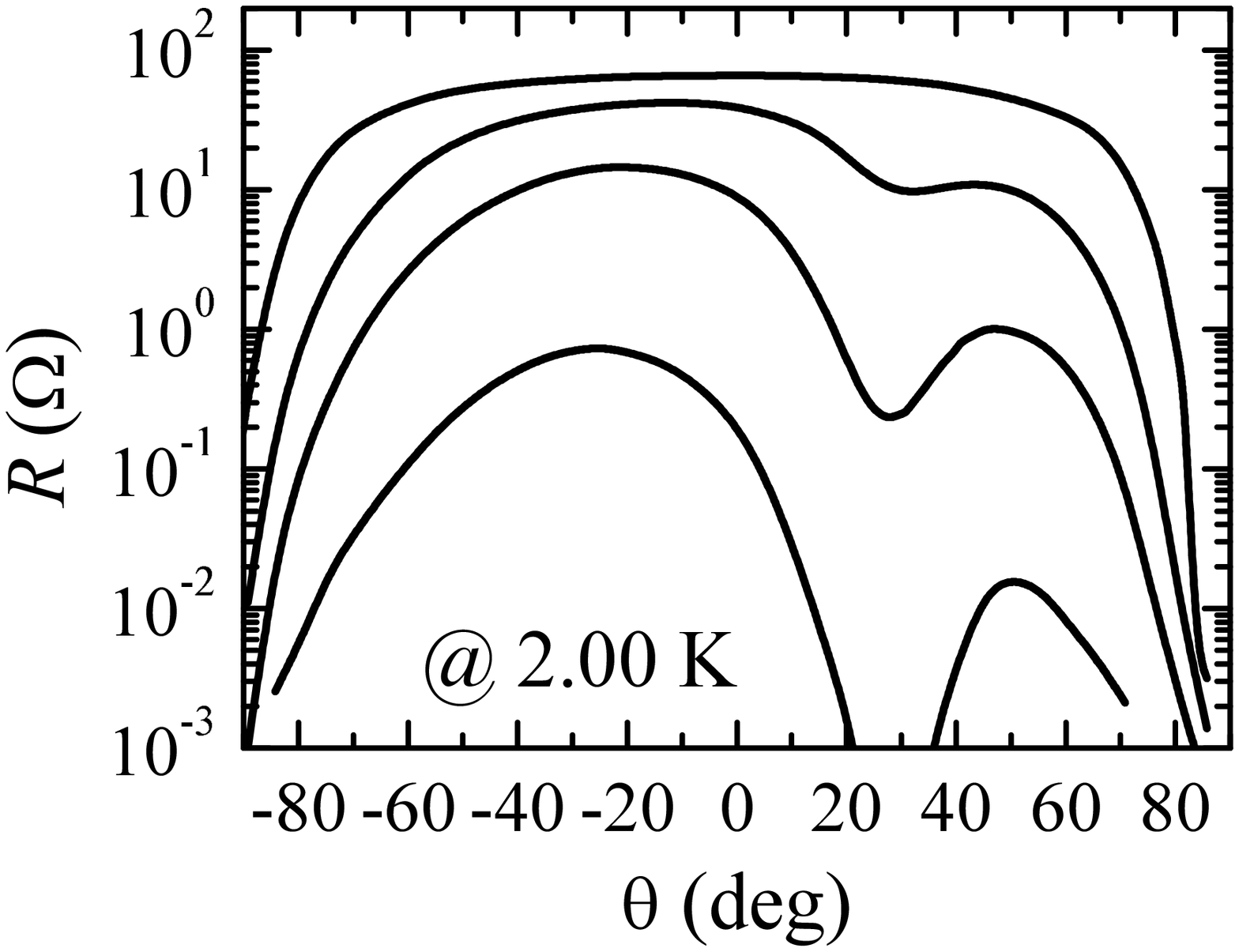}
\caption{Resistance versus magnetic field orientation for sample J40, all curves taken at $T=2.00$~K. The fields are from bottom to top $H=0.2, 0.3, 0.4,$ and $0.6$~T. Intrinsic pinning by the layering reduces the resistance below the measurement limit at $90\grad$ and $-90\grad$ and near $\theta=0\grad$ the resistance is at its maximum. For a temperature-dependent field range a second, local resistance minimum appears around $30\grad$ caused by the pinning of the coplanar defect structure. \label{figure1}}
\end{figure}

Careful determination of the C-direction is very important for the interpretation of the subsequent conductivity measurements. Since the samples had to be broken to obtain cross-sectional views from the TEM investigations which were available only after the completion of the conductivity measurements, the columnar orientation was determined from resistance measurements as a function of $\theta$, where $\theta$ is the angle enclosed between the magnetic field direction and the film normal. For the right combination of temperature and applied field $R(\theta)$-curves show a pronounced dip near the columnar angle in addition to the anisotropy due to the layered structure of the films. Figure \ref{figure1} %
shows a set of $R(\theta)$-curves for sample J40 taken at constant temperature and varying magnetic fields. The C-directions for all the samples were determined from such $R(\theta)$-curves as the angle of the local minimum in resistance; results are given in Table~\ref{table1}. Comparing these angles to those derived from TEM images, a systematic deviation towards larger angles is observed. This trend is also seen in other samples not presented here. The deviation is most likely due to the development of vortex staircases because of the competing pinning by the coplanar defects and the intrinsic pinning by the insulating germanium layers. Studies on Y123 with competing extended pinning sites \cite{Silhanek99} have shown that such staircase structures can experience strong pinning.

Zero field transition temperatures $T_{c}(0)$ and upper critical fields $H_{c2}(T)$ were determined from measurements of the resistance $R$ versus temperature $T$ in applied fields ranging from zero to $2.5$~T. The zero-field data could be well described using the expression for fluctuation conductivity in three dimensional (3D) superconductors. \cite{Schmidt68,Maki68,Thompson70} In-field $R(T)$ data were analyzed according to the scaling approach of Ullah and Dorsey \cite{Ullah90,Ullah91} and assuming the upper critical field to be linear in $T$. The upper critical field was determined for fields in the C and AC-direction, and the extrapolated values at zero temperature were identical in the two orientations within the accuracy of the scaling analysis. The extrapolated $H_{c2}(0)$ values were used to estimate the coherence length $\xi_{ab}(0)$ according to the following relation from Ginzburg-Landau (GL) theory: \cite{GL50}
\begin{equation}%
H_{c2\perp} = \frac{\Phi_{0}}{2\pi\xi_{ab}^{2}},
\end{equation}%
where $H_{c2\perp}$ is the upper critical field perpendicular to the $ab$-planes, $\Phi_{0}$ is the flux quantum $h/2e$, and $\xi_{ab}$ is the coherence length within the $ab$-planes. For moderate anisotropy\footnote{The anisotropy ratio $\gamma=H_{c2\perp}/H_{c2\parallel}$ was not determined for these films, but estimated from previous studies to be less than $1/10$.} the systematic error in $\xi_{ab}$ as a result of using $H_{c2}$ in the C-direction amounts to $<5$\%. \cite{Blatter92} The penetration length $\lambda_{ab}(0)$ at zero temperature can be calculated using the relation of Kes and Tsuei:\cite{Kes83}
\begin{equation}%
\lambda_{ab}(0) = 1.05 \times 10^{-3}\left[\rho_{N}(0)/T_{c}(0)\right]^{1/2}
\end{equation}%
with $\rho_{N}(0)$ the extrapolated normal state resistivity at zero temperature. Values for these important parameters are given in Table~\ref{table2}. It is worth noting the exceptionally large GL parameters $\kappa\equiv\lambda/ \xi \gtrsim 200$ for these multilayered films.
\begin{table}
\caption{Summary of the superconducting parameters for all three samples. Given are the critical temperature $T_{c}(0)$ [K], followed by the upper critical field $H_{c2}(0)$ [T]. The forth and fifth column are the coherence length $\xi_{ab}$ [nm] and the penetration depth $\lambda_{ab}$ [$\mu$m] at zero temperature, respectively. Noted in the last column are the GL-parameters $\kappa = \lambda/\xi$. \label{table2}}
\begin{ruledtabular}
\begin{tabular}{lrrrrr}
 & \multicolumn{1}{c}{$T_{c}(0)$} & \multicolumn{1}{c}{$H_{c2}(0)$} & \multicolumn{1}{c}{$\xi_{ab}(0)$} & \multicolumn{1}{c}{$\lambda_{ab}(0)$} & \multicolumn{1}{c}{$\kappa$} \\\hline
C40 & $2.40$ & $7.4$ & $6.7$ & $1.43$ & $213$ \\
J40 & $2.21$ & $7.0$ & $6.9$ & $1.62$ & $235$ \\
P40 & $2.27$ & $7.3$ & $6.7$ & $1.67$ & $249$ \\
\end{tabular}
\end{ruledtabular}
\end{table}

\section{Activation Energy for TAFF}\label{taff}

It is generally accepted that immediately below $T_{c}(H)$ the vortex system is in a liquid state with no long range order and vanishing shear modulus. The liquid state can be further divided into a free flux flow region, which is extremely narrow for the superconductors under investigation, and a thermally activated flux flow (TAFF) region. The pinning potential which hinders the free flow of vortices is caused by a combination of point-like defects\footnote{caused mostly by oxygen impurities} and the correlated defects of the columnar microstructure. In general, the temperature dependence of the resistance can be described by the following exponential for activated behavior
\begin{equation} \label{eq.activation}%
R \propto \exp\left(\frac{U_{A}(H,T)}{k_{B}T}\right),
\end{equation}%
with $k_{B}$ the Boltzmann constant and $U_{A}(H,T)$ the field and temperature dependent activation energy.

The temperature and field dependence of the activation energy for plastic motion of the flux lines has been calculated within the limits of weak collective pinning \cite{Geshkenbein89,Vinokur90,Vinokur91}
\begin{equation} \label{eq.sqrt}%
U_{\textrm{pl}} \propto \frac{T_{c}-T}{\sqrt{H}}.
\end{equation}%
Replacing $U_{A}$ in Eq.~\ref{eq.activation} with $U_{\textrm{pl}}$ the $T$ and $H$-dependence can be separated and the activation energy as determined from Arrhenius-plots of $R(T)$-curves taken at constant $H$ should follow a simple $H^{-1/2}$ power law. In two-dimensional (2D) or quasi-2D superconductors another mechanism for TAFF is possible, the movement of unbound dislocations. The expression for the activation energy is given as \cite{Feigelman90}
\begin{equation} \label{eq.dislocation}%
U_{\textrm{dis}} \propto \left(T_{c}-T\right) \ln\left(\frac{H_{0}}{H}\right),
\end{equation}%
with $H_{0}$ a characteristic field of the order of $H_{c2}(0)$. This kind of dissipation mechanism is topologically forbidden in three-dimensional superconductors since only dislocations with zero net flux can be realized. The $T$ and $H$-dependence can be separated again when inserting expression (\ref{eq.dislocation}) for $U_{A}$ in Eq.\ \ref{eq.activation} and in thin films or strongly layered superconductors the activation energy for TAFF should obey a logarithmic field dependence in case $U_{\textrm{dis}}<U_{\textrm{pl}}$.

\subsection{Strongly Coupled Layers}

\begin{figure}
\includegraphics[width=85mm,keepaspectratio]{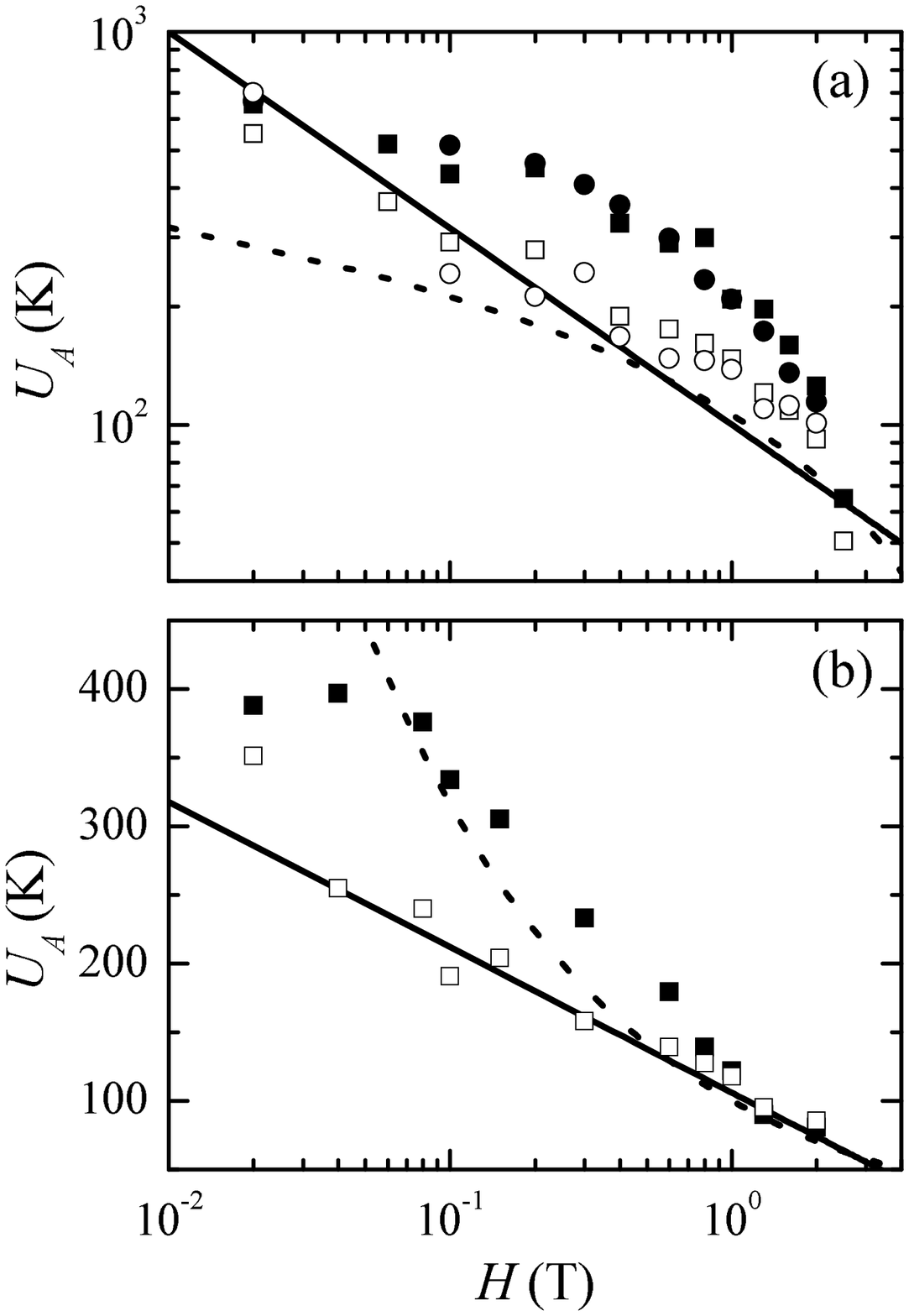}
\caption{Activation energies for TAFF as a function of the external magnetic field. The graph (a) is a double-logarithmic plot and shows results for the strongly coupled samples C40 (squares) and J40 (circles) and C (filled symbols) and AC field orientation (open symbols), respectively. The solid line is the result from a homogeneous reference film \cite{Ruck98} and the dotted line is the $\ln H$ line from graph (b). The results for the more weakly coupled sample P40 are shown in graph (b) in a semi-logarithmic plot, again with both field orientations, C (filled squares) and AC (open squares). The solid line is a least-squares fit of Eq.~\ref{eq.dislocation} to the data from AC-oriented magnetic field. For comparison the dotted line is the same line as the solid line in graph (a). \label{figure2}}
\end{figure}

Figure~\ref{figure2}(a) shows the activation energies in the two strongly coupled samples in a double-logarithmic plot. %
As a guideline, the solid line represents the results for a homogeneous, unstructured Ta$_{\, \textrm{x}}$Ge$_{\, 1-\textrm{x}}$-film fitted to a field-dependence according to Eq.~\ref{eq.sqrt}. \cite{Ruck98} For comparison the $\ln H$ least-square fit in the lower graph (b) is shown as the dotted curve. For C-aligned fields in the range from $0.1$ up to about $2.0$~T both samples represented in (a) have a significantly increased activation energy for TAFF due to increased pinning by extended defects. The enhanced pinning is seen to slowly diminish for fields exceeding $0.2$ to $0.3$~T. If this is identified with the matching field, for which the density of vortices is equal to the number of pinning defects, this corresponds to an average spacing of strong pinning defects of approximately $100$~nm. The $1/\sqrt{H}$-dependence is approximately preserved for fields in the AC-direction, although the activation energies for $H>0.2$~T are consistently higher than for the reference film. We understand this as a combined effect of the layering and the coplanar defect structure in the present films. \cite{Engel01,Engel01b} Overall, the field-dependence of the activation energies of the strongly-coupled films are consistent with anisotropic, three dimensional superconductors.

\subsection{Weakly Coupled Layers}\label{taffweak}

In contrast with the results for the strongly coupled films described in the previous paragraph, the weakly coupled film P40 is more complex. The data for the AC field direction, aligned far from the columns, follows the 2D prediction of Eq.~\ref{eq.dislocation}. To emphasize this behavior the activation energies for this film are displayed in a linear-log plot in Fig.~\ref{figure2}(b). For comparison the $1/\sqrt{H}$-relation from graph (a) is plotted as the curved, dotted line in this semi-logarithmic plot. Especially at lower field values the measured activation energies deviate from the power-law behavior, but can be fitted very successfully using an expression according to Eq.~\ref{eq.dislocation} with a logarithmic dependence on the applied magnetic field (solid line). The fitting parameter, the characteristic field $H_{0}=10.0$~T, is comparable to the upper critical field $H_{c2}(0)=7.3$~T as required by the theory. Thus, the most likely process leading to dissipation is the movement of flux-carrying dislocations, possible only in 2D or quasi-2D, layered superconductors.

The interpretation of the activation energies for fields aligned with the extended defect structure in sample P40 is more difficult. Firstly, it has to be noted that activation energies for fields smaller than $0.8$~T are significantly higher compared to AC-oriented fields. At higher fields on the other hand, the activation energies are equal for the two field orientations within experimental errors, in contrast to the strongly coupled samples where the difference between the orientations remains significant up to at least $2.0$~T. Although the field-dependence can be roughly described by the power-law behavior of Eq.~\ref{eq.sqrt}, with increased $U_{A}$ over a certain field-range similar to C40 and J40, the values obtained below $0.1$~T are significantly below the expectations of Eq.~\ref{eq.sqrt} and approach the results from AC-oriented fields, just as at high fields.

It appears unlikely that the differences between the C and AC aligned fields can be understood based on strictly uncoupled quasi-2D layers, for in this picture the extended defects act as simple point defects centered on the superconducting layers. We propose instead a picture involving dimensional transitions, suggested by theoretical descriptions of pancake liquids \cite{Bulaevskii98} and by experiments on Bi2212 containing columnar defects. \cite{Sato97,Kosugi97,Morozov99} At small magnetic fields and corresponding high temperatures pancake vortices are trapped by the extended defects, but independently in each layer. As the field is increased the correlations in the direction of the applied field increase and the pancakes recouple to form vortex lines. In this intermediate field range the activation energies for TAFF are much higher for C than for AC-aligned fields. At even higher magnetic fields the magnetic intra-layer interactions begin to dominate and the pancakes decouple again.

\section{Current-Voltage Characteristics} \label{ivcurves}

The state and dynamics of the vortex fluid can be studied directly by measuring the current-voltage relation in samples exposed to a magnetic field. The interpretation of these data are based on the recognition that the product of current and field determines the force applied to the vortices, while the measured voltage is proportional to the product of the field and the vortex flow velocity. Thus the usual presentation of the data, as a plot of voltage versus current, is equivalent to a plot of the speed of the response (flow velocity) vs.\ force for the vortex fluid. In this section we describe $I$-$V$ results for a range of field and temperatures, probing the vortex dynamics over the portion of the phase diagram accessible within the limits of our experimental arrangements.

One of the most fundamental issues, whether a vortex state with true zero resistance exists for $T>0$ in the mixed phase of dirty type-II superconductors, is still unresolved. \cite{Strachan01,Landau02} The simple question whether the barriers for flux creep diverge in the limit of zero applied current below a field dependent transition temperature is inherently difficult to answer experimentally. The most popular theoretical model predicting a second order phase transition from a vortex liquid into a low-temperature solid vortex phase with zero resistance is based on an analogy to magnetic spin glass systems. \cite{Huse92} Depending on the anisotropy of the superconductor and the character of the defects (point-like or extended) one differentiates between 2D and 3D vortex glasses (VG) \cite{Fisher89,Fisher91} and Bose-glasses (BG). \cite{Nelson92}

At the second order phase transition a characteristic length and corresponding time diverge algebraically and these define a set of critical glass exponents. A scaling ansatz between electric field $E$ and current density $J$ is obtained from which a number of relations between applied current $I$ and voltage $V$ can be derived for both the vortex liquid and solid state. Those relations which were important for the analysis of the $IV$-characteristics presented here will be repeated. Neglecting some specific differences between the VG and BG theories these relations can be summarized conveniently using a parameter $D$, with $D=2, 3$ for quasi-2D and 3D VG and $D=4$ for BG. Right at the glass transition temperature $T=T_{g}$ the voltage increases as a simple power-law of the applied current
\begin{equation} \label{power-law}
V \propto I^{\left(z+1\right)/\left(D-1\right)}.
\end{equation}
As the transition is approached from above ($T>T_{g}$) the resistivity $\rho$ vanishes as
\begin{equation} \label{resistivity}
\rho \propto |T-T_{g}|^{\nu\left(z+2-D\right)},
\end{equation}
and below the transition temperature $IV$-curves show an exponential relation with zero linear resistance
\begin{equation} \label{exponential}
V \propto \exp\left[-c\left(I_{c}/I\right)^{\mu}\right].
\end{equation}
Relations (\ref{resistivity}) and (\ref{exponential}) cross over to critical power-law behavior above a crossover current given by
\begin{equation} \label{crossover}
I_{\textsf{x}} \propto |T-T_{g}|^{\nu\left(D-1\right)}.
\end{equation}
$\nu$ and $z$ are the static and dynamic critical exponents, respectively; they should be field-independent and equal for films which belong to the same universality class. $\mu$ is the glass exponent and will be discussed in detail below, $c$ and $I_{c}$ are material-dependent parameters. Appropriately rescaled sets of $IV$-curves should collapse onto just two universal functions above and below the transition temperature using the critical exponents $\nu$ and $z$ as obtained from above relations. In order to use this condition as an additional check of the analysis $(V/I)|T-T_{g}|^{-\nu(z-D+2)}$ is plotted versus $I|T-T_{g}|^{\nu(1-D)}$.

However, the collapse of $IV$-curves onto an upper and lower branch as the only criterion for a vortex glass transition is not suitable. Due to the experimentally limited current and voltage resolution a collapse may be achieved for a range of glass exponents. Voss-de Haan \emph{et al.} \cite{Voss99,Voss00} have demonstrated the collapse of a set of $IV$-curves for deliberately limited current ranges but varying parameters. It is therefore mandatory to achieve a collapse of the $IV$-curves and to fulfill Eq.~\ref{power-law} - \ref{crossover} simultaneously for the same set of parameters.

\subsection{$IV$-Curves in Strongly Coupled Layers}

\begin{figure}
\includegraphics[angle=-90,width=85mm,keepaspectratio]{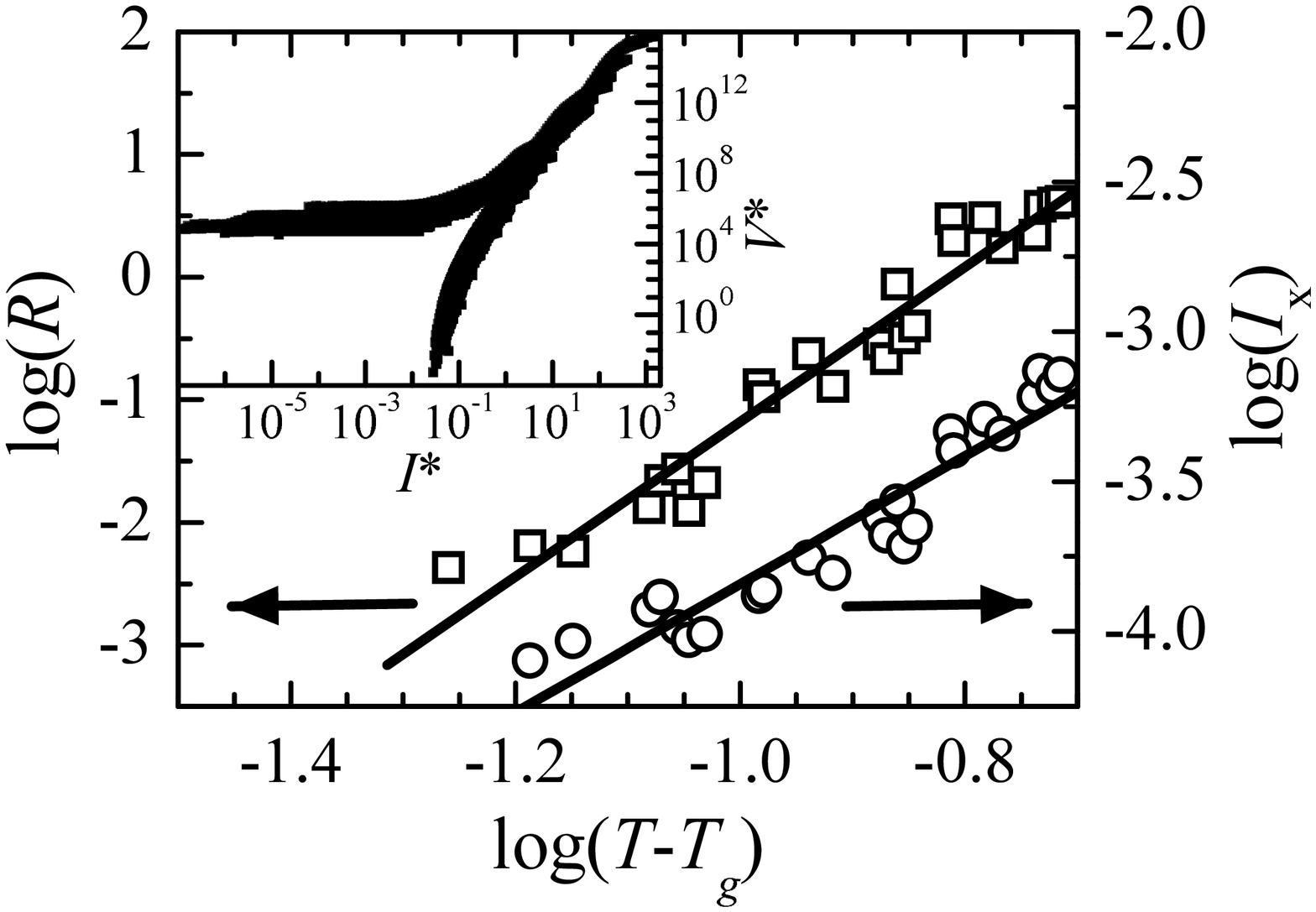}
\caption{Scaling analysis of $IV$-curves for sample J40 and magnetic fields $H=0.1,\ 0.2,\ 0.3,\ 0.4,\ 0.5,\ 0.6,\ 0.8,\ 1.0,$ and $1.3$~T applied in the AC-direction. The main graph shows a double-logarithmic plot of the linear resistance $R$ above $T_{g}$ (left axis) and the crossover current $I_{\textsf{x}}$ separating linear from nonlinear behavior versus the reduced temperature $T-T_{g}$. Linear least-square fits to the data allow the determination of the exponents in Eqs.~\ref{resistivity} and \ref{crossover}. Assuming a 3D-vortex glass ($D=3$) the glass exponents $\nu$ and $z$ can be calculated and the inset shows the resulting collapse of the $IV$-curves in a double-logarithmic representation using these values. Here $I^{\ast}=I|T-T_{g}|^{\nu(1-D)}$ and $V^{\ast}=(V/I)|T-T_{g}|^{-\nu(z-D+2)}$ are the rescaled current and voltage. \label{figure3}}
\end{figure}

As an example, Fig.~\ref{figure3} shows the analysis of $IV$-curves taken for sample J40 and nine magnetic field magnitudes in the range $0.1$ to $1.3$~T applied in the AC direction at constant temperature. %
The temperature range spans typically about $300$~mK on either side of the glass transition temperature $T_{g}$ in intervals of $50$~mK. The main graph is a double-logarithmic plot of the linear resistance $R$ at low currents above $T_{g}$ (left ordinate) and the crossover current $I_{\textsf{x}}$ (right ordinate) marking the onset of nonlinear behavior in the vortex liquid phase above $\Tg$ versus $T-T_{g}$. $I_{\textsf{x}}$ was defined here as the current for which the resistance exceeded twice the linear resistance. To construct the plot the value of $\Tg$ was first estimated as the temperature which showed power-law behavior (Eq.~\ref{power-law}). Subsequently these $T_{g}$ values were adjusted until the data in Fig.~\ref{figure3} followed straight lines. A least-squares fit to the data allows one to determine the exponents in Eqs.~\ref{resistivity} and \ref{crossover}.

Knowing these exponents one can check the collapse of the $IV$-curves as shown in the inset of Fig.~\ref{figure3} in a double-logarithmic plot of the rescaled current $I^{\ast}=I|T-T_{g}|^{\nu(1-D)}$ and voltage $V^{\ast}=(V/I)|T-T_{g}|^{-\nu(z-D+2)}$. Often this procedure already leads to a very satisfactory collapse. If not, the parameters are slightly adjusted and then the scaling of the linear resistance and crossover current checked and if necessary corrected, and so forth until all conditions for a glass transition are fulfilled within the experimental limits. To extract the glass exponents $\nu$ and $z$ one needs to know which of the different glass models is appropriate, i.e.\ $D=2, 3$ or $4$. As discussed in an earlier paper \cite{Engel01} we use $D=3$ for the strongly coupled samples C40 and J40 for AC-aligned fields and the BG model, $D=4$, when the external fields were applied in the C-direction. The resulting critical glass exponents listed in Table~\ref{table3} are very similar to those reported for vortex glasses in HTSCs. \cite{Grigera98,Voss99,Ammor00,Klein00} We stress that our values are field-independent over at least one order in magnitude of the applied field.
\begin{table}
\caption{Glass scaling parameters for all three samples and for magnetic fields applied in the C-direction (BG model) and AC-direction (3D-VG model). $\nu$ are the static and $z$ the dynamic glass exponents. Also given are parameters characterizing the glass melting line in the phase diagrams Figs.~\ref{figure6} and \ref{figure7}, whereby $\chi$ is the disorder parameter according to Eq.~\ref{Bosemeltingline} and $\nu_{0}$ the glass exponent of Eq.~\ref{3Dmeltingline}. The weakly-coupled sample P40 showed no clear glass transition for AC-aligned fields. \label{table3}}
\begin{ruledtabular}
\begin{tabular}{ldddddd}
 & \multicolumn{3}{c}{Bose-glass (C)} & \multicolumn{3}{c}{vortex-glass (AC)}\\
 & \multicolumn{1}{c}{$\nu$} & \multicolumn{1}{c}{$z$} & \multicolumn{1}{c}{$\chi$} & \multicolumn{1}{c}{$\nu$} & \multicolumn{1}{c}{$z$} & \multicolumn{1}{c}{$\nu_{0}$} \\\hline
C40 & $1.0$ & $8.7$ & $0.50$ & $1.4$ & $5.1$ & $1.0$ \\
J40 & $0.7$ & $8.7$ & $0.38$ & $1.1$ & $6.9$ & $1.0$ \\
P40 & $1.3$ & $7.0$ & $0.64$ & -- & -- & -- \\
\end{tabular}
\end{ruledtabular}
\end{table}

Strachan \emph{et al.}\ argue that the above analysis is still not sufficient evidence for a transition to a vortex state with vanishing linear resistance. \cite{Strachan01} According to Eqs.~\ref{power-law} - \ref{crossover} $IV$-curves plotted in double-logarithmic fashion must change their curvature from positive above $T_{g}$ to negative below the transition temperature. This property is preserved in the universal functions, which are specified by the scaled data sets in the inset of Fig.~\ref{figure3}. Whereas individual data sets above $\Tg$ describe the upper branch including the characteristic kink to the horizontal part, rescaled current-voltage curves below $\Tg$ describe only very short sections and only a set of $IV$-curves for different $T-\Tg$ give a full representation of the lower branch as in the inset. This problem arises because of the extremely rapidly decreasing voltage with decreasing current and the experimental voltage resolution allows $IV$-curves to be taken over only a relatively small current range. As a consequence an apparent collapse of $IV$-curves can be achieved for a range of scaling parameters as reported by Voss-de Haan \cite{Voss99,Voss00} and Strachan \emph{et al.}\cite{Strachan01} demand the detection of positive curvatures above the glass transition and negative curvatures below as a necessary condition for a vortex glass transition. 

Due to the extremely rapidly vanishing voltage level at low temperatures for decreasing currents, our data do not fulfill this experimentally extremely demanding condition for a second order phase transition. However, data taken for the weakly coupled film includes $IV$-characteristics at constant field which are obviously lacking any sign change in curvature as will be demonstrated below. Based on these observations and thermodynamic measurements on Y123 [Ref.~(\onlinecite{Roulin98})] we believe that an interpretation of our data as a second order phase transition into a vortex glass at low temperatures is justified. 

\subsection{$IV$-Curves in Weakly Coupled Layers}\label{ivweak}

\begin{figure}
\includegraphics[width=85mm,keepaspectratio]{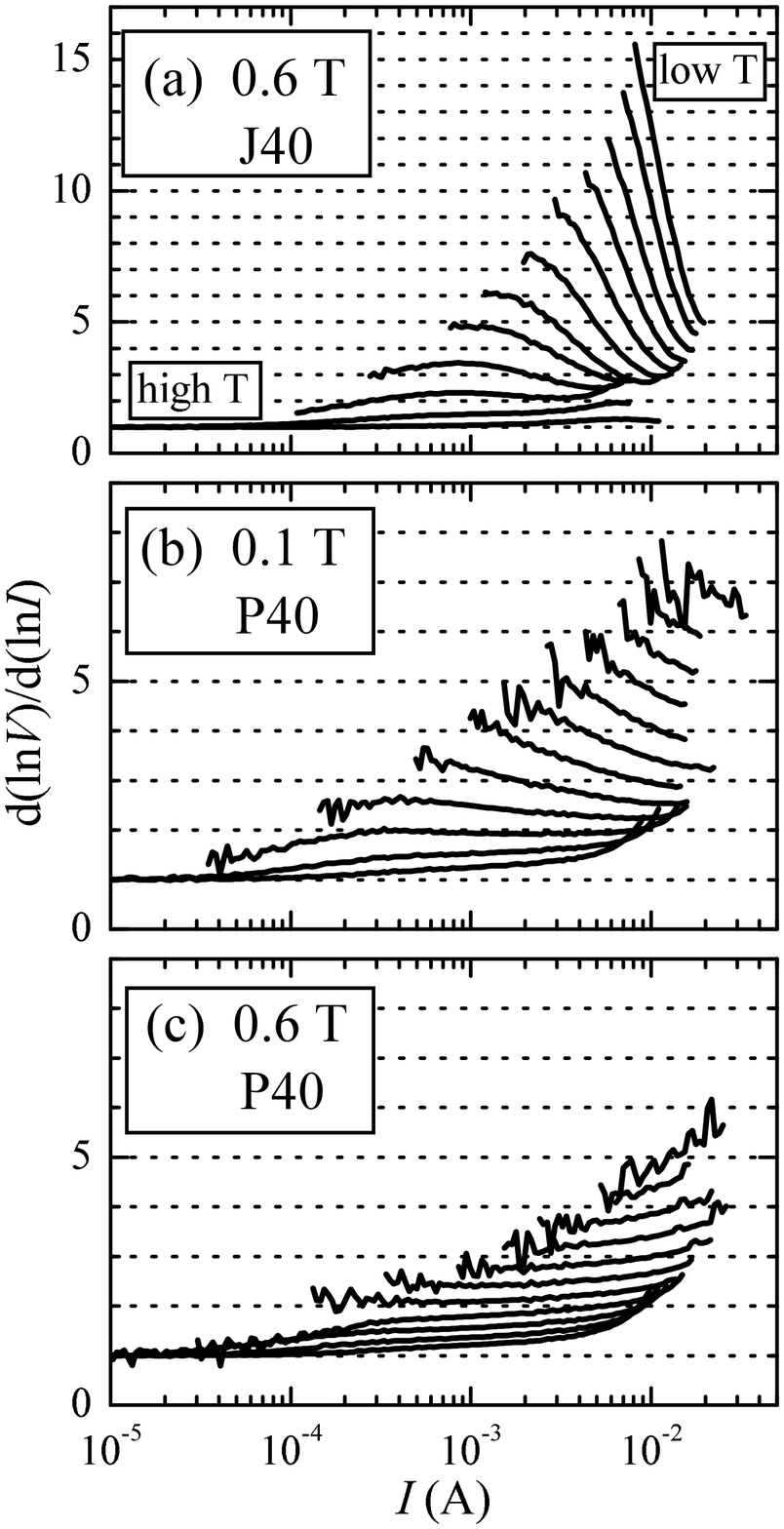}
\caption{Logarithmic gradient $\textrm{d}(\ln V)/\textrm{d}(\ln I)$ versus applied current $I$ at magnetic fields $H=0.1$ (b) and $0.6$~T (c) applied in the C-direction for sample P40. For comparison we show in (a) similar data from a strongly coupled film (J40, $0.6$~T, C). The temperatures range from $1.40$ to $1.95$~K (a), $1.60$ to $2.15$~K (b) and $1.40$ to $1.90$~K (c) in steps of $50$~mK. At the lower magnetic field (b) and for the strongly coupled sample (a) the curves' gradients change from positive at high temperatures to negative at low temperatures; this translates to a change in curvature of the original $IV$-curves in a double-logarithmic plot as required by glass theories. At the higher magnetic field (c) curves with negative gradient are absent. \label{figure4}}
\end{figure}

The $IV$-curves measured for the weakly coupled film P40 differ qualitatively from those of the strongly coupled films. The differences are most easily seen by plotting the logarithmic gradient $\textrm{d}(\ln V)/\textrm{d}(\ln I)$ versus the applied current, as displayed in Fig.~\ref{figure4} for a selection of situations. %
For reference, Fig.~\ref{figure4}(a) includes data from a strongly coupled film showing the high-temperature nearly-ohmic behavior with a logarithmic derivative very close to unity for all currents. As the temperature is lowered the derivative is close to unity (ohmic) at low and high currents, passing through a maximum (super-ohmic) at intermediate currents. At even lower temperatures, below the glass transition temperature, the low-current ohmic behavior is missing; the continuing rise of the logarithmic derivative with decreasing current signals a voltage which falls more rapidly than any power law with decreasing current. Note that this presentation identifies immediately the problem of using $IV$-curves to establish the presence of super-ohmic behavior at vanishing currents as a signature of a solid vortex phase. At low currents, where the higher temperature curves show an unambiguous tendency toward ohmic behavior, the voltage level at lower temperatures is too small to be measured.

We now compare this behavior to that of film P40, shown for C-oriented fields in Figs.~\ref{figure4}(b) and (c). At the lowest applied field, Fig.~\ref{figure4}(b), the logarithmic derivatives are qualitatively similar to those in a strongly coupled film. Their behavior at lower temperatures continues to rise with decreasing current, with no suggestion of a low-current ohmic dependence in the accessible current range. In a higher field, however, that behavior is entirely different, with even the low-temperature curves showing a tendency to fall at the lowest currents. There is clearly no pinned solid vortex state at all in the weakly coupled film at these higher C-oriented fields. Interestingly, similar data for a field of $0.3$~T shows nearly horizontal lines at all temperatures, corresponding to the power-law behavior expected at a phase transition.

Further insight into the vortex states in this film is gained by application of the $IV$ scaling analysis. At the low fields of Fig.~\ref{figure4}(b) the analysis is successful, but at higher fields only the upper branch ($T>\Tg$) can be formed. That high-field scaling analysis returns a transition temperature within our measurement range. The lower temperatures cannot be accommodated as in the glass phase, nor can they be accommodated within a model in which the glass transition is assumed to be at temperatures below our lower limit. It appears that a \emph{would-be} transition is sensed in the vortex liquid phase, but that the transition never, in fact, takes place.
 
\begin{figure}
\includegraphics[angle=-90,width=85mm,keepaspectratio]{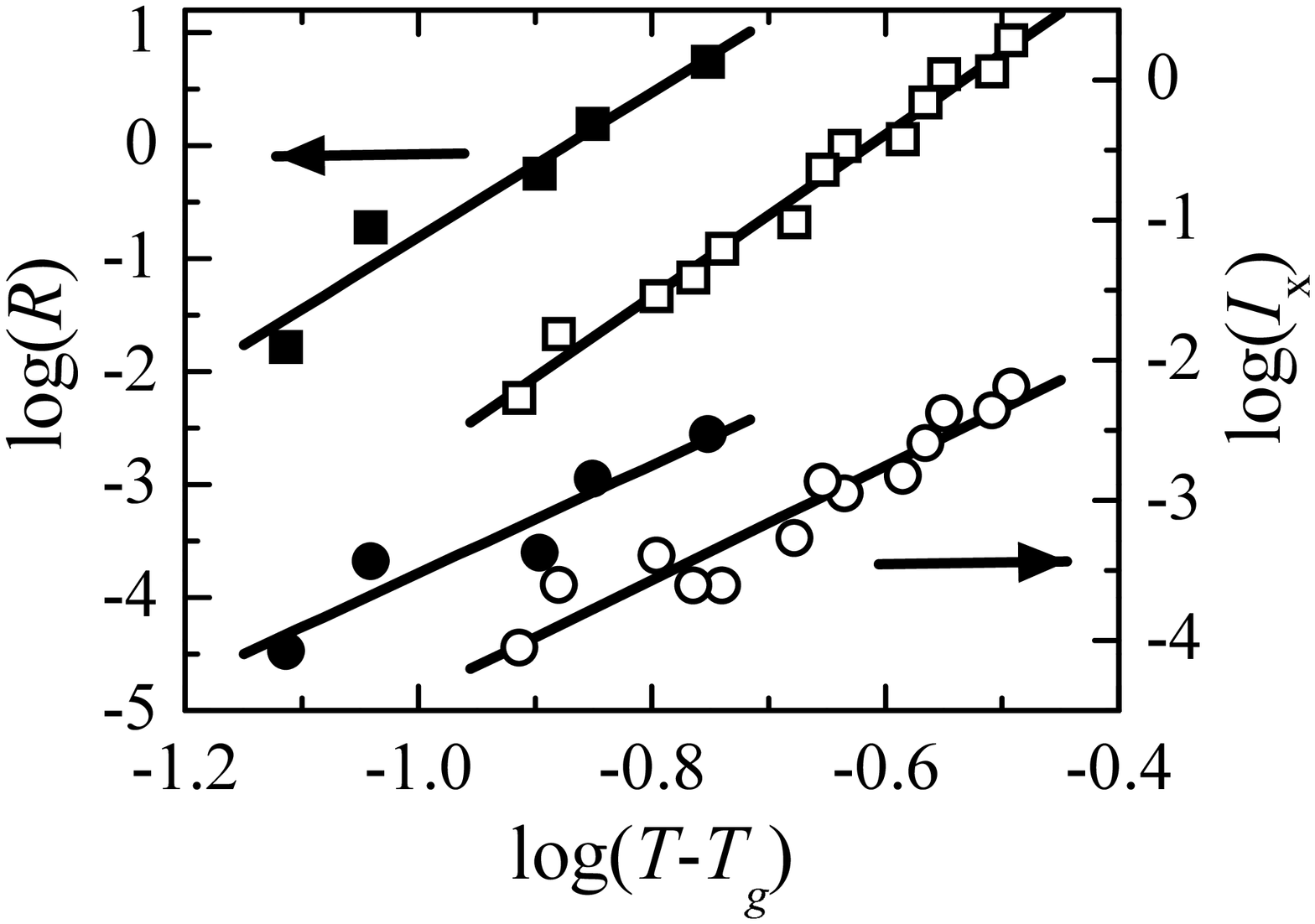}
\caption{Scaling analysis of the high temperature $IV$-curves for P40 at magnetic fields from $H=0.1$ to $1.3$~T. The linear resistance (left axis, squares) and the crossover current (right axis, circles) are plotted double-logarithmically versus $T-\Tg (H)$. The data clearly separate into two groups for fields $H=0.1$ and $0.3$~T (filled symbols) and those at $H \geq 0.6$~T (open symbols). \label{figure5}}
\end{figure}

The scaling analysis suggested by Eqs.~\ref{resistivity} and \ref{crossover} is still available for these results, for it relies on data only above the glass transition. The resulting scaling of $R$ and $I_{\textsf{x}}$ is shown in Fig.~\ref{figure5} analogous to the main graph in Fig.~\ref{figure3}. %
But contrarily to the scaling of J40 and the other strongly coupled samples for which all data points followed the same power-law behavior irrespective of the applied field strength, the data for P40 clearly separate into a low field ($H=0.1$ and $0.3$~T, filled symbols) and a high field group ($H \geq 0.6$~T, open symbols). Assuming BG behavior at low fields, the $IV$-curves collapse very satisfactorily onto the two universal functions above and below the transition temperature and with comparable glass exponents as the BG phases in C40 and J40. The $IV$-curves at higher fields and temperatures above the estimated transition temperature collapse as well. Not surprisingly, the collapse at lower temperatures ($T<\Tg (H)$) is not possible, because the $IV$-curves have still positive curvature. It is significant that a lower assumed transition temperature, placing the $IV$-curves at the lowest temperatures above the transition, leads to a violation of the collapse and scaling of $R$ and $I_{\textsf{x}}$. Thus in a C-oriented field of $0.6$~T the behavior above the transition temperature determined from scaling of $R$ and $I_{\textsf{x}}$ is exactly as expected for a liquid, but below that temperature the vortex fluid resembles neither a liquid nor a glass. Similar behavior is observed also for all investigated fields higher than 0.6~T.

Turning the magnetic field in the AC-direction so that field and defect structure are misaligned leads to the disappearance of even the low-field glass phase; at the lowest field $H=0.1$~T the $IV$-curves show only positive curvatures. Consequently, an attempt to analyze the data according to glass theories leads to the same difficulties as for the high field $IV$-curves and C-aligned fields. 

The TAFF activation energies in Section~\ref{taff} suggest an interpretation for the lack of a clear second-order liquid-glass transition in this film. Recall that those energies had a field dependence characteristic of moving dislocations in two-dimensional superconductors for AC alignment, and that the C-aligned energies converged with those 2D values at fields above $0.6$~T. These observations suggest that the low-temperature vortex fluid state is essentially 2D for AC-aligned fields, while in C-aligned fields there is a crossover from a 3D glass phase at low fields to a mobile 2D phase at higher fields.

This conclusion relates directly to the qualitative change seen in the $IV$-characteristics. At low fields, the pancake vortices are highly correlated along the defect structure and at low temperatures they freeze into a Bose-glass phase. With increasing magnetic field the interlayer correlations are lost and the superconducting layers become effectively decoupled, preventing the formation of a BG phase as has been suggested also to occur in ion-irradiated Bi2212. \cite{Morozov99} For misaligned magnetic fields (AC), on the other hand, the interlayer correlations between pancake vortices are not strong enough to form a 3D-VG for any applied field.

The true nature of the low-temperature, high-field phase remains unclear, however. For a truly 2D superconductor Eqs.~\ref{power-law}
-- \ref{crossover} are not valid, instead it is expected that the glass phase exists at $T=0$~K, only. \cite{Fisher91} At finite temperatures $IV$-curves should still be divided into a low-current linear and a high-current nonlinear part. The crossover current relation Eq.~\ref{crossover} is replaced by
\begin{equation} \label{2Dcrossover}
I_{\textsf{x}}^{2D} \propto T^{1+\nu_{2D}},
\end{equation}
with $\nu_{2D}=2$. Analyzing the $IV$-characteristics of P40 according to Eq.~\ref{2Dcrossover} over the accessible temperature range leads to a non-constant exponent with $\nu_{2D}\gtrsim 30$, most certainly ruling out a truly 2D-VG phase. 

\section{Vortex Phase Diagrams} \label{phasediagrams}

\subsection{Strongly Coupled Layers}

\begin{figure}
\includegraphics[angle=-90,width=85mm,keepaspectratio]{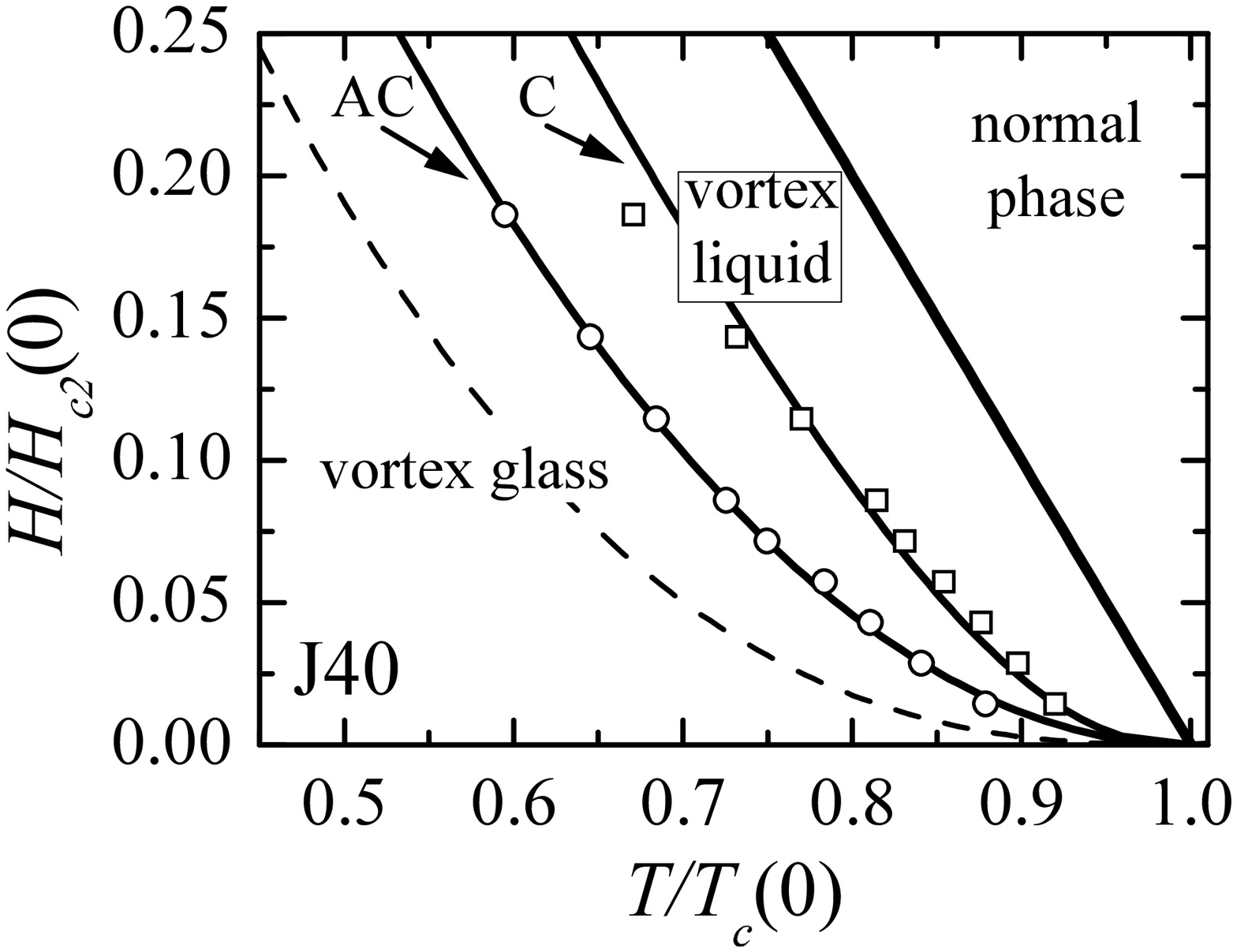}
\caption{Reduced $H$-$T$ vortex phase diagram for the strongly coupled sample J40. The uppermost straight line is the upper critical field $H_{c2}(T)$. The experimental glass transition temperatures are given as squares (C-aligned fields) and circles (AC). The glass melting temperatures were fitted to expressions as described in the text (solid curves). To demonstrate the effects of the coplanar defect structure the glass melting line for a homogeneous Ta$_{0.27}$Ge$_{0.73}$-film without correlated defects \cite{Ruck98} is given as a reference (dashed line). The correlated defects shift the melting line to significantly higher temperatures, whereby the shift for C-aligned fields is approximately twice the shift for AC-aligned fields. The vortex phase diagram for sample C40 is almost identical. The glass melting line in the AC-direction is the same, only for C-aligned fields the melting line is at somewhat lower temperatures (about $5$\% lower reduced temperature for reduced fields in the range $0.05$ to $0.15$). \label{figure6}}
\end{figure}

The analysis in the previous section puts us into a position to plot vortex phase diagrams for each of the three samples. Turning first to the strongly-coupled samples Fig.~\ref{figure6} shows the reduced $H$-$T$ phase diagram for J40. %
The pattern is almost identical to that in C40. The uppermost line is the linearized upper critical field $H_{c2}(T)$ separating the normal from the superconducting phases. The vortex glass melting temperatures as derived from $IV$-curves are plotted as squares (C-) and circles (AC-alignment). At lower fields and temperatures the vortices form the respective Bose-glass and 3D-VG for C and AC-aligned fields. In the 3D-VG model the glass melting line can be described as a simple power-law \cite{Fisher91}
\begin{equation} \label{3Dmeltingline}
[T_{c}(0)-\Tg] \propto H^{1/2\nu_{0}}
\end{equation}
with an expected $\nu_{0}\simeq 2/3$. The experimentally determined $T_{g}(H)$ do follow a power-law behavior very nicely, but with an exponent $\nu_{0}=1$. 

The field dependence of the BG melting line for C-aligned magnetic fields is more complicated. Blatter \emph{et al.}\cite{Blatter92} give an expression derived for superconductors containing columnar defects and applied magnetic fields smaller than the matching field, but large enough that the vortices interact, $B>\Phi_{0}/\lambda^{2}$:
\begin{equation} \label{Bosemeltingline}
\Tg(B) \approx \chi T_{m}(B) + (1-\chi)T_{c}\left(1-\frac{B}{\mu_{0}H_{c2}(0)}\right),
\end{equation}
with $T_{m}(B)$ the melting temperature of the flux line lattice (FLL) for the material in the clean limit without columnar or point defects. $\chi$ is a disorder parameter which equals unity in the clean limit and approaches zero for very strong disorder. The clean-limit melting temperature is unknown for these \TaGex-films. Instead, we used the glass melting line for a homogeneous Ta$_{0.27}$Ge$_{0.73}$-film without extended defects but with weak-pinning point defects \cite{Ruck98} (dashed curve in Fig.~\ref{figure6}) to approximate $T_{m}$ in the above equation. A fit of Eq.~\ref{Bosemeltingline} then provides a satisfactory description of the field-dependence of the Bose-glass melting line and the disorder parameter $\chi$ a quantitative measure of the shift of the glass transition to higher temperatures.

This phenomenological fitting procedure yields the disorder parameters listed in Table~\ref{table3}, from which it can be seen that the disorder appears marginally stronger in film J40 than in C40. We have not been able to correlate this difference with any measured structural parameters, though it is worth noting that the superconducting layers are thinner in the more strongly disordered film.

\subsection{Weakly Coupled Layers}

\begin{figure}
\includegraphics[angle=-90,width=85mm,keepaspectratio]{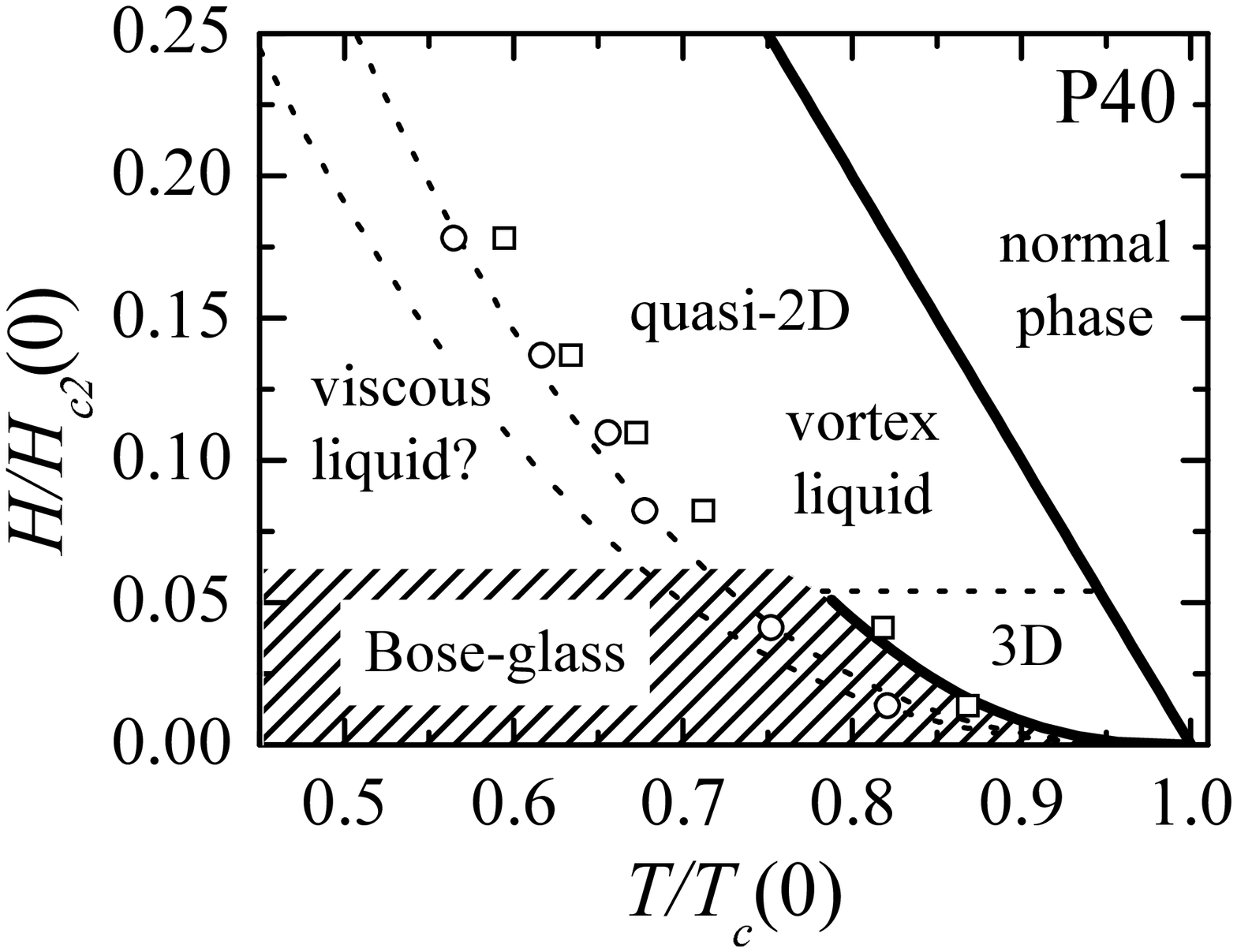}
\caption{Reduced $H$-$T$ vortex phase diagram of sample P40 analogous to Fig.~\ref{figure6}. The solid straight line is again the upper critical field and the lowest dotted curve is the reference glass melting line of a Ta$_{0.27}$Ge$_{0.73}$-film. For C-aligned fields a Bose-glass phase could only be found at low applied fields (hatched area) separated from a 3D vortex liquid by the BG melting line (solid curve) according to Eq.~\ref{Bosemeltingline} fitted to the two available data points (two lowest squares). As described in Sec.~\ref{ivcurves} at higher fields there were no indications for a glassy phase. The squares at higher fields mark the temperature for which a transition to a zero-resistance state is expected from the extrapolation of resistance data in the quasi-2D liquid phase. The dashed line is a guide to eye, only. In the AC case (circles) this high-field scenario extends to the lowest measured fields, the dotted line is a power-law fit to the data points.\label{figure7}}
\end{figure}

The phase diagram for the weakly coupled film P40 is dominated by its quasi-2D behavior and a solid vortex glass phase is only realized for small, C-oriented fields. In Fig.~\ref{figure7} the reduced $H$-$T$ phase diagram is shown analogous to Fig.~\ref{figure6}. %
The uppermost solid line is the upper critical field $H_{c2}(T)$ and the lowest dotted curve is the glass melting line of the reference Ta$_{0.27}$Ge$_{0.73}$-film. \cite{Ruck98}

As outlined in Sec.~\ref{ivweak} Bose-glass scaling procedures could only be successfully applied to $IV$-curves taken at the two lowest fields applied parallel to the defect structure. The derived glass melting temperatures $\Tg$ are given as the two lowest squares in Fig.~\ref{figure7}. A glass melting line according to Eq.~\ref{Bosemeltingline} has been successfully fitted to these data points (solid curve); the resulting disorder parameter indicates reduced disorder compared to the strongly coupled samples (see Table~\ref{table3}). At temperatures below $\Tg$ the vortex system forms a solid BG (hatched area), which melts into a 3D vortex liquid at higher temperatures. The three dimensional character of the vortex liquid in this field range is supported by the field-dependence of the activation energies for TAFF (Sec.~\ref{taffweak}). Both 3D-phases are limited by an upper magnetic field, signaled in the case of the BG by the failing low temperature scaling and in the case of the vortex liquid deduced from activation energies. The approximate crossover to 2D behavior is given by the horizontal dotted line. It is worth mentioning that we have found signs of an reentrant 2D vortex liquid in Sec.~\ref{taffweak} at very low magnetic fields, not shown in Fig.~\ref{figure7}.

For higher magnetic fields and all field values applied in the AC-direction the $IV$-characteristics gave no evidence of a pinned solid vortex phase. The would-be transition temperatures sensed by the high-temperature vortex liquid are given as squares (C) and circles (AC). The dashed and dotted lines are guides to the eye and may mark the crossover to a highly viscous vortex liquid. This remains however an open question which demands further investigations.

A comparison of these phase diagrams with vortex phase diagrams for HTSCs is also interesting. The strongly coupled samples resemble those for weakly anisotropic Y123 [Ref.~(\onlinecite{Krusin94,Nakielski96})] and isotropic (K,Ba)BiO$_{3}$. \cite{Klein00} P40 on the other hand shows remarkable similarities with the moderately anisotropic Bi$_{2}$Sr$_{2}$Ca$_{2}$Cu$_{3}$O$_{10}$ [Ref.~(\onlinecite{Li94})] and especially Tl$_{2}$Ba$_{2}$Ca$_{2}$Cu$_{3}$O$_{10}$. \cite{Yu97}

\section{Dynamics in the Bose-glass Phase} \label{dynamics}

The exponential $IV$-characteristics in the glass phase, Eq.~\ref{exponential}, are caused by vortex creep due to finite applied currents and finite probabilities for vortex excitations at non-zero temperatures. These vortex dynamics in the glass phase can be characterized by the glass exponent $\mu$, with $\mu > 0$ for a truly glassy phase. Various models for vortex creep in the presence of defects allow the calculation of the field and temperature-dependence of $\mu$. For weak pinning by point defects the weak collective pinning (WCP) theory offers one way to calculate the glass exponent $\mu$. However, $\mu$ depends non-trivially on the various length scales determining the size of the vortex bundles, which are the smallest independently moving objects in WCP theory. \cite{Feigelman89,Nattermann90,Blatter94}

Within the Bose-glass theory for superconductors containing extended defects vortex creep is explained by the excitation of vortex loops, half-loops and kinks. \cite{Nelson93} Under the acting Lorentz-force due to the finite probing current, these excitations lead to the characteristic creep of the vortices through the pinning potential. These ideas can be developed for the case of pinning by columnar defects \cite{Nelson93} as well as for pinning by coplanar defects. \cite{Marchetti94,Marchetti95} The problem can be mapped onto electron transport in doped semiconductors \cite{Mott69,Shklovskii84} wherefore it is also known as variable range hopping (VRH) of vortices. The following results are obtained for VRH by the creation of double superkinks
\begin{equation} \label{VRH1}
\mu = 1/3\ \textrm{(columnar defects)},
\end{equation}
\begin{equation} \label{VRH2}
\mu = 1/2\ \textrm{(coplanar defects)}.
\end{equation}
And for larger currents half-loop excitations can extend under the acting Lorentz-force until they reach the next favorable pinning site leading to a glass exponent
\begin{equation} \label{halfloop}
\mu = 1.
\end{equation}
Several groups have found some evidence for VRH in Y123 [Ref.~(\onlinecite{Dekker92a,Thompson97})] and Bi2212. \cite{Soret99,Soret00,Ammor00}

To check the applicability of these results to our films it is necessary to extract the exponent $\mu$ from $IV$-curves measured at temperatures below the glass melting temperature $\Tg$. After recasting Eq.~\ref{exponential} into
\begin{equation} \label{exponent}
\frac{\textrm{d}\left(\ln V/I\right)}{\textrm{d}I} \propto I^{-\left(\mu+1\right)}
\end{equation}
this is most easily done by plotting the left side of Eq.~\ref{exponent} versus the current $I$ on a double-logarithmic scale. The data should then follow a straight line with gradient $-(\mu+1)$ from which $\mu$ can be easily calculated. We apply this analysis in the following two sections.

\subsection{Weakly Coupled Layers}

\begin{figure}
\includegraphics[angle=-90,width=85mm,keepaspectratio]{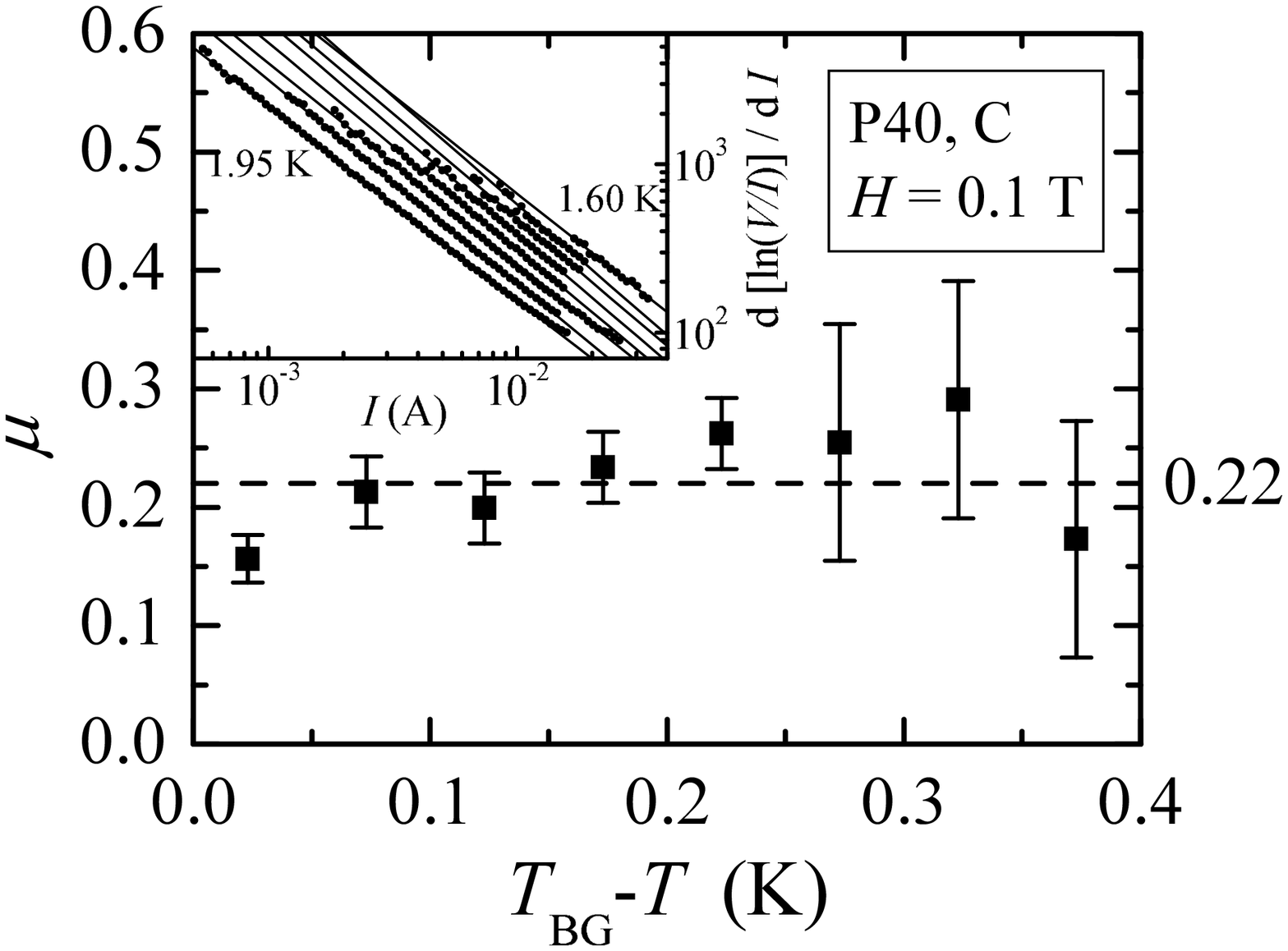}
\caption{Glass exponent $\mu$ versus $\Tg-T$ for P40 and $H=0.1$~T applied in the C-direction. The exponent shows no significant temperature dependence over the measured $T$-range with an averaged $\overline{\mu}=0.22$. The inset shows the $IV$-data recalculated according to Eq.~\ref{exponent} in a double-logarithmic plot and least-squares fits to the data from which the exponent was extracted. The high current part of the $IV$-data where they deviate from the exponential behavior is truncated. \label{figure8}}
\end{figure}

We start the discussion of the experimental results with the weakly coupled sample P40. From the phase diagram Fig.~\ref{figure7} it can be seen that a solid glass phase exists only for small, C-oriented fields. In Fig.~\ref{figure8} the glass exponent $\mu$ is plotted as a function of the temperature below the glass melting temperature, $\Tg-T$, for $H=0.1$~T. %
The inset shows the respective rescaled $IV$-curves and least square fits according to Eq.~\ref{exponent}, demonstrating the applicability of above model to our data. Within the estimated errors the glass exponent $\mu$ is temperature independent with an averaged $\overline{\mu}=0.22$. This is approximately half the theoretically calculated value of $1/2$ for VRH in the presence of coplanar pinning sites (Eq.~\ref{VRH2}) and still significantly less than the other predicted value for pinning by columnar defects (Eq.~\ref{VRH1}).

Although above $IV$-characteristics can be described successfully with an exponential dependence according to Eq.~\ref{exponential}, existing models describing the dynamics in the Bose-glass phase fail to explain the magnitude of the glass exponent. The available data did not allow for a conclusive determination of a field-dependence of $\mu$. The only other set of $IV$-curves taken in the Bose-glass phase at $H=0.3$~T is very close to the upper field limit of the glass phase. These $IV$-curves are very close to power-law behavior resulting in glass exponents $\mu \approx 0$.

\subsection{Strongly Coupled Layers}

\begin{figure}
\includegraphics[angle=-90,width=85mm,keepaspectratio]{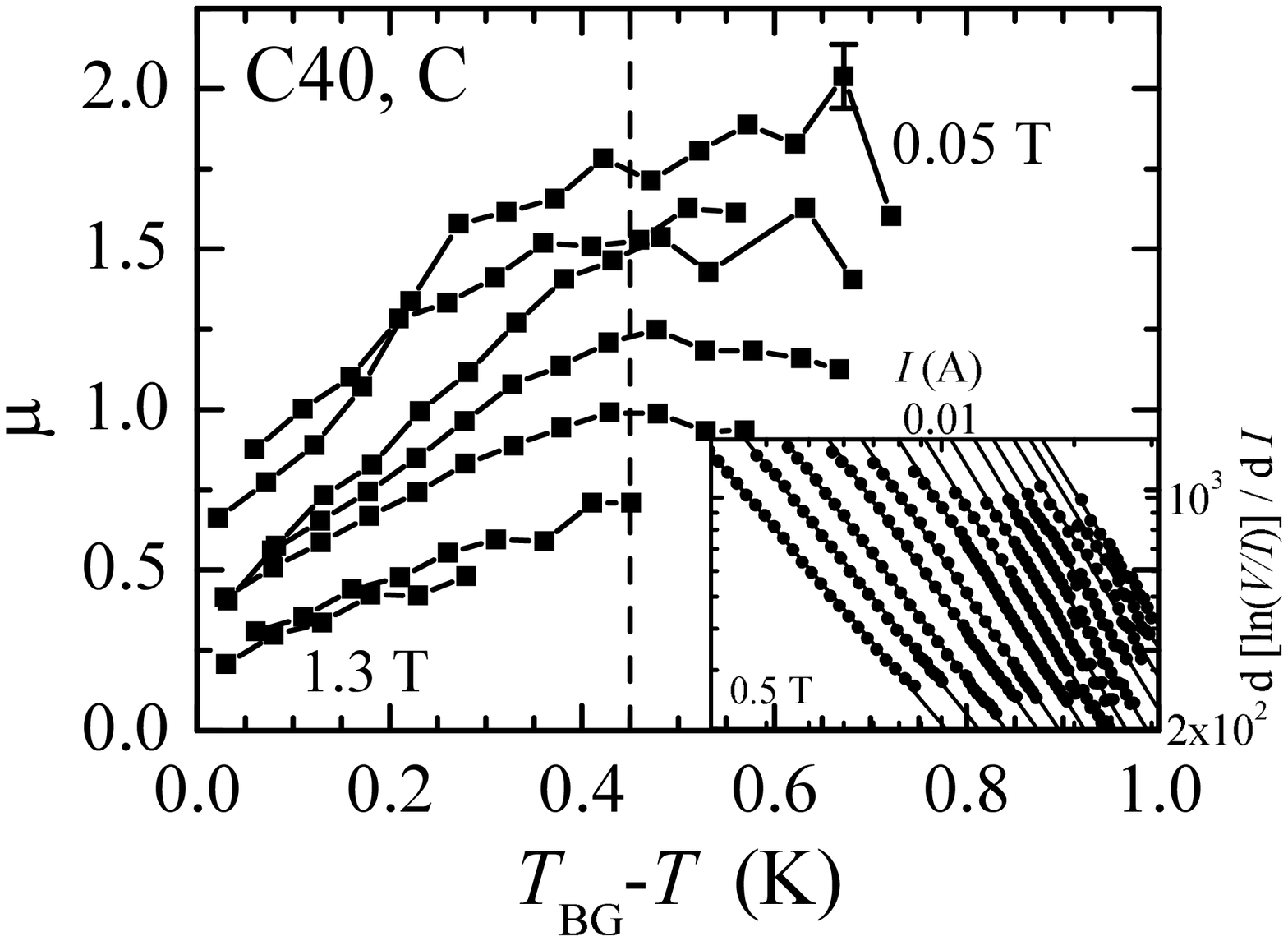}
\caption{Temperature and field-dependence of the glass exponent $\mu$ for C40 and C-aligned magnetic fields. Field magnitudes range from $0.05$ T at the top over $0.1,\ 0.3,\ 0.5,\ 0.7,\ 1.0$ to $1.3$ T at the bottom. At fixed temperatures below the glass melting temperature $\Tg$ the glass exponent $\mu$ decreases monotonically with increasing magnetic field. At fixed field magnitudes first the exponent rises when the temperature is lowered below $\Tg$ until it levels off at a field-dependent value. The crossover to constant $\mu$ happens at field-independent $\approx 0.45$ K below $\Tg$, marked by the vertical dashed line. This crossover is also visualized in the inset, where the rescaled data and least square fits are plotted analogous to the inset in Fig.~\ref{figure8}. Typical uncertainties in $\mu$ are given as the error bar for the top most data point. \label{figure9}}
\end{figure}

The glass exponent $\mu$ has a markedly different temperature-dependence in the strongly coupled samples C40 and J40, and the extended field-range of the glass phases additionally enabled us to study the field-dependence. In Fig.~\ref{figure9} $\mu$ is plotted versus $\Tg-T$ for a wide range of applied fields and temperatures below $\Tg(H)$ in the case of C40 and C-oriented fields. %
Very similar behavior was observed for magnetic fields applied in the AC-direction and the other strongly coupled film J40.

Looking at the field-dependence for fixed $\Tg-T$ the glass exponent $\mu$ decreases monotonically with increasing field. This is true for the complete observed temperature and field range; the latter spans more than one order in magnitude. The temperature-dependence shows even more interesting characteristics. As the temperature is lowered below the glass transition temperature $\Tg$ the glass exponent increases down to temperatures approximately $0.45$~K below $\Tg$. This increase may well be described by a linear dependence on temperature. Extrapolating $\mu$ to $\Tg$ one obtains non-zero glass exponents at the glass transition, which decrease with increasing field as at lower temperatures. The rate at which $\mu$ increases also decreases with higher magnetic field. It is important to note that this happens in a temperature interval below $\Tg$ for which we observed a constant $\mu$ in the weakly coupled sample P40 (compare Fig.~\ref{figure8}).

At a temperature roughly $0.45$~K below $\Tg$ (marked by the dashed vertical line in Fig.~\ref{figure9}) a crossover to an approximately constant glass exponent is observed. \footnote{For the two highest magnetic fields the temperature range was limited by our $1.2$~K limit. Therefore, the crossover could not be observed.} The level is again field-dependent and ranges from about $1.75$ at low fields to estimated $0.5$ at the highest fields. The correlation between the crossover to a constant $\mu$ and $\Tg-T$ links the glass exponent to the formation of the glass phase, but is not explained by the simple VRH picture presented above.

\section{Conclusions}\label{conclusions}

We have performed extensive conductance measurements on artificially layered \TaGex multilayers and compared the results for films with strong and weak interlayer coupling and with and without extended strongly pinning defects, respectively. Due to their large GL parameter $\kappa$ conduction below $T_{c}(H)$ is dominated by the dynamic behavior of vortices in response to an applied transport current and the interactions between vortices and the samples' microstructure.

The strongly coupled films are best modelled as weakly anisotropic three dimensional superconductors. The activation energies for TAFF show a significant increase above the expected $1/\sqrt{H}$-dependence when the magnetic field is aligned with the extended pinning sites. The $IV$-characteristics were interpreted within the existing vortex glass models. A thorough analysis results in a very consistent second order phase transition from a vortex liquid at high temperatures to a pinned vortex glass phase at a field-dependent transition temperature $\Tg(H)$. The critical glass exponents $\nu$ and $z$ are field-independent and all $IV$-curves taken at different fields and temperatures could be collapsed onto two scaling functions. The correlated defects lead to a large shift of the glass transition line to higher reduced temperatures when compared to an unstructured homogeneous film. This shift is maximized for fields well aligned with the correlated defect structure.

In the case of the more weakly coupled sample P40 one has to distinguish between fields parallel to the pinning sites (C-oriented) and magnetic fields at large angles to the columnar structure (AC-oriented). For the latter situation the film behaves like an anisotropic quasi-2D superconductor. The activation energies for TAFF are best described by a $\ln H$-dependence over the complete magnetic field range investigated, typical for 2D superconductors. Additionally, we have seen no real signs of a second order phase transition into a glass phase at low temperatures, in line with predictions that a glass phase in two dimensions should at best exist at $T=0$. 
On the other hand, turning the magnetic field into the C-direction leads to a reoccurrence of 3D vortex behavior in a narrow field range. The strongest argument for this interpretation is the formation of a BG phase at low magnetic fields. This three dimensional vortex phase is supported by the extended defects which assist the formation of well-aligned stacks of pancake vortices leading to increased correlations across the superconducting layers as suggested by results on Bi2212. \cite{Sato97,Kosugi97,Morozov99}
This picture is further supported by the qualitative change observed in the $IV$-characteristics between high and low fields.

Although the low-temperature $IV$-curves can be described very consistently within the BG model there are significant differences in the vortex creep between the strongly and weakly coupled samples and neither of them is fully explained by the simple VRH mechanism. This is particularly true for vortex creep in the strongly coupled samples. Since the glass exponent $\mu$ showed very similar $T$ and $H$-dependence for C and AC-aligned fields, details of the pinning sites, such as point-like or extended, seem to be less important and the magnitude of $\mu$ varied over a wide range with temperature and field. However, the strong correlation between the crossover from increasing $\mu$ to a constant value and the temperature difference $\Tg-T$ suggests a connection between the formation of the glass phase and the dynamics of the vortices. More specific, two fundamental variables of the glass theories, the correlation length $\xi$ and the relaxation time $\tau$, are proportional to $|\Tg-T|^{x}$, where $x$ is determined by the critical exponents $\nu$ and $z$. To elucidate a possible connection between the $T$ and $H$-dependence of $\mu$ on the one hand and $\nu$ and $z$ on the other, more theoretical work in this area is called upon.

The authors acknowledge support by the New Zealand Marsden Fund and thank J.\ C.\ Abele for assistance in gathering experimental data.

\bibliography{bibsc}

\end{document}